\documentclass[doublespacing]{elsart}
\usepackage{amssymb}
\begin{document}
\begin{frontmatter}
\title{CPN Models in General Coordinates}
\author{K.J.Barnes}
\address{Department of Physics and Astronomy, University
of Southampton, Highfield, Southampton, SO17 1BJ, United Kingdom.}
\begin{abstract}
An analysis of CPN models is given in terms of general coordinates
or arbitrary interpolating fields. Only closed expressions made
from simple functions are involved. Special attention is given to
CP2 and CP4.  In the first of these the retrieval of stereographic
coordinates reveals the hermitian form of the metric.  A similar
analysis for the latter case allows comparison with the
Fubini-Study metric.
\end{abstract}

\begin{keyword}
Kahler \sep stereographic \sep complex projective

\PACS \sep 11.10 Lm \sep 11.25.-W \sep 11.30. Pb \sep 11.30. Rd
\end{keyword}
\end{frontmatter}
\noindent{\bf Introduction}

Despite the central importance of CPN models [1], and their recent
revival as arising in supersymmetric from from minimized linear
models [2] following the revision of the underlying supersymmetry
algebra of densities to include central terms, there appears to be
no treatment of them in general coordinates which would allow
arbitrary field redefinitions for the interpolating Goldstone
Bosons.  In this article just such an analysis is presented.  The
next section explains how this is achieved by embedding the
necessary structure into a more complicated one.  Strangely,
perhaps, nothing is needed but simple functions, and a completely
general solution is found in closed form.  In the following
section the special case of CP2 and stereographic coordinates is
presented.  Then the corresponding step is made for CP4 allowing
the connection to the Fubini-Study metric.  Finally, there are
brief conclusions and suggestions are made for future work.
\newline \hrule width 14cm

{\bf General Framework}

\noindent{} Curiously this section begins by consideration of the
embedding of the structure needed for the current problem into
that of a larger system which has previously been solved in
general coordinates leading to a closed form involving only simple
functions [3].  The embedding is unique.  Thus the starting point
is a review of this established larger system and its solution, in
which the liberty of changing notation (slightly) for convenience
has been taken.

Consider then the Lie algebra of $SU_n$ specified by taking as a
basis the set of $(n^2-1)$ traceless hermitian $n\times n$
matrices $\lambda_i$ with the product law
\begin{equation}
 \lambda_i\lambda_j = {2\over n} \delta_{ij} + (d_{ijk} + if_{ijk})
\lambda_k \end{equation}
\noindent as specified by Gell Mann [4].
When specific values of the structure constants $f_{ijk}$, or the
symmetric $d_{ijk}$ tensors or the $\lambda_i$ matrices themselves
are needed then the notation of reference [4] will be assumed.  An
element of the group $SU_n$, $g(\theta)$ is specified in
exponential form by a set of $(n^2-1)$ real parameters $\theta_i$,
so that in infinitesimal form the transformations
\begin{equation}
q_A \longrightarrow q_A - {i\over 2} \theta^i(\lambda_i)_{AB} q_B
\end {equation}
and
\begin{equation} M_i \longrightarrow M_i + \theta_k f_{ikj}M_k
\end {equation}
specify the behaviour of the basic spinor $q_A$ (quark) and
adjoint vector $M_i$ fields.  Now define a traceless matrix M by
\begin{equation}
M_{KL} = M_i (\lambda_i)_{KL}
\end {equation}
so that
\begin{equation}
M_i = {1 \over 2}Tr.(M \lambda_i) \end{equation}

\noindent then a group element $g(\theta)$ which induces a unitary
transformation
\begin{equation} q_A \longrightarrow U(\theta)_{AB}q_B \end {equation}
on the basic spinors clearly induces an orthogonal transformation
\begin{equation} M_i \longrightarrow R_{ij}M_j = {1 \over 2}Tr.(U^{-1}\lambda_iU
\lambda_j)M_j\end{equation} on the adjoint representation.

The algebra of $SU_n \times SU_n$ is spanned by two sets of
$(n^2-1)$ orthogonal elements $L_i$ and $R_i$ satisfying the
commutation relations
\begin{equation}
[L_i,L_j] = if_{ijk}L_k
\end{equation}
\begin{equation}
[R_i,R_j] = if_{ijk}L_k \end{equation}
\begin{equation}
[L_i,R_j] = 0 \end{equation}

\noindent and the linear combinations
\begin{equation} V_i = L_i + R_i \end{equation}
\begin{equation} A_i = L_i - R_i \end{equation}
are frequently used.  Obviously the $V_i$ generate a $SU_n$
subgroup which is parity conserving.  An element of the $SU_n
\times SU_n$ group may be specified by two sets of $(n^2-1)$ real
parameters, and the alternative expressions
\begin{equation} g = exp(-i[\theta^V_iV_i +
\theta^A_iA_i]) \end{equation}
\noindent and
\begin{equation} g = exp(-i\theta^L_iL_i)
exp(-i\theta^R_iR_i)\end{equation}
\noindent will prove useful
with
\begin{equation}
\theta^L_i = \theta^V_i + \theta^A_i
\end{equation}

\begin{equation}
\theta^R_i = \theta^V_i - \theta^A_i
\end{equation}

\noindent specifying the correspondence.  Every element of the
group can also be decomposed into a product of the form
\begin{equation}
g = exp(-i\phi_iA^i) exp(-i\theta_iV^i)
\end{equation}

\noindent which is unique in a neighbourhood of the identity
element and this will play a crucial role in the general nonlinear
realization scheme.  The linear transformation laws are best
specified by giving the quarks a Dirac spinor index in the usual
manner and taking
\begin{equation}
q \longrightarrow q - {i\over 2} \theta^L_i \lambda_i
{(1+\gamma_5)\over 2} q - {i\over 2} \theta^R_i \lambda^i
{(1-\gamma_5)\over 2} q
\end{equation}

\noindent as the concrete infinitesimal form.

\noindent Since the matrices
\begin{equation}
P_L = {(1 + \gamma_5)\over 2}
\end{equation}

\noindent and
\begin{equation}
P_R = {(1 - \gamma_5)\over 2}
\end{equation}

\noindent act as a standard set of projection operators, the
treatment of linearly transforming multiplets of $SU_n \times
SU_n$ now follows trivially.

To treat the nonlinear realizations of $SU_n \times SU_n$ in full
generality the $(n^2 - 1)$ hermitian components $M_i$ of the
adjoint vector of $SU_n$ must be considered in more detail.  In
the terminology of Michel and Radicati [5], the vector is said to
be generic (or to belong to the generic stratum) if all
eigenvalues of $M$ are distinct.  For the generic case the minimal
polynomial for the matrix is the characteristic polynomial
satisfying the equation
\begin{equation} \prod^n_{A=1} (M - m_A) = 0 \end{equation}
where the $m_A$ are the eigenvalues which satisfy
\begin{equation} \sum^n_{m=1} m_A = 0 \end{equation}
if the matrix is traceless.  Thus the (n - 1) vectors with
components given by powers of the matrix in the form
\begin{equation} {M^\alpha}_{i} = {1\over 2} Tr.([M]^\alpha
\lambda_i)\hskip 2cm [\alpha = 1,2, ..., (n-1)] \end{equation}

\noindent are a linearly independent set, and the quantities
\begin{equation} T_A = Tr.([M]^A) = \sum^n_{B=1}[m_B]^A \equiv
\sum^n_{B=1} m_{AB} \end{equation}

\noindent are $(n - 1)$ independent $SU_n$ invariants. ($S_1$ is
identically zero.)  At once it is clear that the general vector
which can be constructed from the $M_i$ has the form
\begin{equation} \xi_i = F_\alpha M_{\alpha i} \end{equation}

\noindent where the $F_\alpha$ are functions of the $(n - 1)$
independent $SU_n$ invariants.  This freedom has been discussed at
length by Gasiorowicz and Geffen [6].  From the point of view of
field theory it corresponds to freedom of choice of interpolating
fields.  Provided that $F_1(0)$ is taken to be unity, and parity
is respected, then all $\xi_i$ so defined are equally good
interpolating fields.  From a geometrical viewpoint the $\xi_i$
may be regarded as coordinates of points of the $(n^2 - 1)$
dimensional coset space manifold formed by the quotient of
$SU_n\times SU_n$ by the vector $SU_n$ subgroup. The freedom is
then viewed as the ability to change coordinates within a local
patch near the origin.

An arbitrary point on the manifold is parameterized by
\begin{equation} exp(-i\xi_i A_i) \equiv L(M) \equiv exp\left( {-i\theta\over 2 \phi}M_i
\lambda_i[P_L - P_R]\right)
\end{equation}

\noindent where the first form corresponds with equation (17) and
the second form represents the appropriate expression when
equation (25) has been used so that the $M_i$ are regarded as the
coordinates, and $\phi^2 = M_iM_i$.

 The general theory is well described by Coleman, Wess and
Zumino [7] and Callan, Coleman, Wess and Zumino [8], and the
geometrical approach by Isham [9].  With the decomposition given
in equation (17) the action of a general element $g$ of the full
group may be written as

\begin{equation} g exp(-i\xi_iA_i) = exp
(-i\xi'_iA_i) exp (-i\eta_iV_i) \end{equation}
\begin{equation} \equiv L(M') exp (-i\eta_i V_i) \end{equation}

\noindent where $M'_i$ and $\eta_i$ both depend on $M_i$ and $g$.
Then the primary result of the general theory is that
\begin{equation} g : M_i \longrightarrow M'_i \end{equation}

\noindent gives a nonlinear realization of the algebra which is
linear on the $SU_n$ vector subgroup.  Moreover if $h$ is an
element of the vector subgroup and
\begin{equation} h : \Psi_\Omega \longrightarrow D(h)_{\Omega\Gamma}
\Psi_\Gamma \end{equation}

\noindent is a linear (unitary) representation of that subgroup,
then
\begin{equation} g : \Psi_\Omega \longrightarrow D[exp(-i\eta_i
\Psi_i)]_{\Omega\Gamma}\Psi_\Gamma \end{equation}

\noindent gives a realization of the full group. Notice that this
latter transformation is linear in $\Psi$ but nonlinear (through
$\eta_i$) in the $M_i$ when $g$ is not in the vector subgroup.
Fields which transform according to equation (31) are called
standard fields, and it is important to understand that by a
suitable redefinition of coordinates any nonlinear realization of
$SU_n \times SU_n$ which is linear on the vector subgroup can be
brought into this standard form.  In practice the most useful
result is that, if one has a linear irreducible (unitary)
representation of $SU_n \times SU_n$ such that

\begin{equation} g : N_\Omega \longrightarrow D[g]_{\Omega \Gamma}
N_\Gamma \end{equation}

\noindent then
\begin{equation} \Psi_\Omega (M) = D
[L^{-1}(M)]_{\Omega \Gamma} N_\Gamma \end{equation} transform as
the components of standard fields.

It is now clear that there are just three classes of fields to
consider:

\begin{enumerate}{\roman{enumi}}

\item Linear representations which may be built up in the usual
way as multispinors with transformation laws defined by equation
(18). These will not be treated in more detail.
\item Vectors $M_i$ transforming as the adjoint
representation of $SU_n$ with a nonlinear transformation law under
chiral action specified by equation (27).  These will allow a
description of the massless Goldstone Bosons (pions etc.)
corresponding to the axial degrees of freedom spontaneously
violated.  The specification of invariants constructed
(nonlinearly) from these is most important and will be exhibited
later.
\item Standard fields which appear linearly in their
transformation laws, but with nonlinear functions of the $M_i$
induced according to equations (31) and (28).  These are important
in describing matter (eg nucleons) interacting with the Goldstone
Bosons as chiral matter.  Once more, the specification of the
corresponding invariants is most important and will be given
later. \end{enumerate}

The technical problem of finding the invariants is solved in
reference [3].  A crucial step is the resolution of the powers of
the matrix $M$ in the form
\begin{equation} [M]^A = [m_B]^AP_B \equiv m_{AB}P_B \end{equation}
where the $P_B$ are $n$ hermitian matrices, each $n \times n$,
with the properties
\begin{equation} P_AP_B = \delta_{AB} P_B \hskip 1cm {\rm (no\hskip 0.2cm sum)} \end{equation}
\begin{equation} Tr.(P_A) = 1 \end{equation} and
\begin{equation} \sum^n_{A=1} P_A = 1 \end{equation}
where this $1$ is the unit $(n \times n)$ matrix.  Although the
$P_A$ are not in general diagonal, the above projection operator
properties make calculations tractable.  Now define
\begin{equation} P_{Ai} = {1 \over 2} Tr.(P_A \lambda_i) \end{equation}
and
\begin{equation} (P_A)_{MN} = P_{Ai} (\lambda_i)_{MN} + {1 \over
n} \delta_{MN} \end{equation} where because the $P_A$ are complete
it follows that
\begin{equation} \sum^n_{A=} P_{Ai} = 0 \end{equation}

\noindent and, introducing
\begin{equation} p_{Ai} = \sqrt 2 [P_{Ai} - (1 + \sqrt
n)^{-1}]P_{ni} \end{equation}

\noindent with
\begin{equation} \sqrt 2 P_{Ai} = p_{Ai} + {1 \over \sqrt n}p_{ni}
\end{equation}

\noindent establishes that $p_{\mu i}$ for $\mu = 1,2, ...,(n -
1)$ are orthonormal.

\noindent The second-rank tensors defined by the $M_i$ are
conveniently handled by an extension of these ideas, and fall into
two classes. One such class is formed by the $n(n - 1)$
independent tensors defined by
\begin{equation} (P_{AB})_{ij} \equiv P_{AiBj} \equiv {1 \over 2}
Tr. (P_A \lambda_i P_B \lambda_j) \hskip 1cm(A \not= B)
\end{equation}

\noindent and
\begin{equation}
I_{ij} = {1 \over 2} Tr. (P_A \lambda_i P_A \lambda_j),
\end{equation}
which have the properties
\begin{equation} II = I \end{equation}
\begin{equation} IP_{AB} = 0 = P_{AB}I \end{equation}
and
\begin{equation} P_{AB}P_{CD} = \delta _{AC} \delta _{BD} P_{AB} \hskip 1cm {\rm (no\hskip 0.2cm sum)}
\end{equation}

\noindent in terms of the matrix notation of the last section.
Moreover, these are all hermitian matrices and the trace of each
$P_{AB}$ is unity.  Since it is easy to show also that
\begin{equation} {\sum_{A\not = B}}'
P_{AB} = 1 - I \end{equation}

\noindent where the sum is over all A and B but excluding terms
with $A = B$, this gives a projection operator resolution in one
sector of the space of these second-rank tensors and so $I$ (with
trace $[n - 1]$) will decompose further.  The second class of
tensors may be identified with the $(n - 1)^2$ independent
matrices with components
\begin{equation} (p_{\alpha \beta})_{ij} \equiv p_{\alpha i} p_{\beta j}
\end{equation}

\noindent which span the subspace of $(n^2 - 1) \times (n^2 - 1)$
matrices projected out on multiplication by $I$ from both sides
and which are therefore orthogonal to the subspace in which the
$P_{AB}$ lie. Since the $p_{\alpha i}$ are orthonormal, the
multiplication law for the $p_{\alpha \beta}$ is
\begin{equation} p_{\alpha \beta} p_{\gamma \delta} = \delta_{\beta
\gamma}p_{\alpha \delta} \end{equation}

\noindent It has been established by Barnes and Delbourgo [10]
that all the independent second-rank tensors which can be
constructed from the $M_i$ are spanned by the $(n - 1)(2n - 1)$
independent $p_{\alpha \beta}$ and $P_{AB}$.

The most general unitary unimodular matrix $U$ constructed from
the $M_i$ may be written in the form
\begin{equation} U = U_A P_A = exp\left[{{-i} \over 2}\theta_A\right]P_A
\hskip 0.4cm {\rm where }\hskip 0.3cm \sum^n_{A=1} \theta_A =
0\end{equation}

\noindent but the $\theta_A$ are otherwise completely arbitrary
independent functions of the independent $SU_n$ invariants $S_A$
subject to the considerations of parity and weak field limits as
mentioned before.  These $(n - 1)$ effective arbitrary functions
of the $(n - 1 )$ invariants are characteristic of the general
solution and will persist throughout this work.

It has been conventional to define
\begin{equation} \sqrt 2 \phi_A = m_A - (1 + \sqrt n)^{-1}m_n \end{equation}
with
\begin{equation} m_A = \sqrt 2 (\phi _A + n^{-{1 \over 2}}\phi_n) \end{equation}
so that, extending the notation used previously,
\begin{equation} M_i = \phi_{\alpha} p_{\alpha i} \end{equation}
and
\begin{equation} \phi_{\alpha, i} = p_{\alpha i} \end{equation}
follow immediately.  Similarly, defining
\begin{equation} \sqrt 2 \psi_A = \theta_A - (1 + n^{1 \over
2})^{-1} \theta_n \end{equation}

\noindent with
\begin{equation} \theta_A = \sqrt 2(\psi_A + n^{-{1 \over 2}}\psi_n) \end{equation}
the $\psi_\alpha$ may be treated as $(n - 1)$ independent
(arbitrary) functions of the $\phi_\alpha$ which then serve as the
$(n - 1)$ independent invariants.

The transformation laws for all realizations are now given in
reference [3] in closed form and in terms of simple functions.
Restricting attention to first-order derivatives of the fields
with respect to space and time, and also restricting attention to
a study of the Goldstone Boson fields $M_i$ and the standard
fields the results can be given in terms of the general analysis
of references [7] and [8].  There are two important results.
First, although $\partial_\mu M_i$ and $\partial_\mu \Psi_\Gamma$
do not transform as standard fields, the covariant derivatives
\begin{equation} D_\mu M_i = A_{\mu i} \end{equation}

\noindent and \begin{equation} D_\mu \Psi_\Gamma = \partial_\mu
\Psi_\Gamma - i \nu_{\mu i} (T_i)_{\Gamma \Omega} \Psi_\Omega
\end{equation}

\noindent where under $SU_n$
\begin{equation} \Psi_\Gamma \longrightarrow \Psi_\Gamma -
i\theta_i (T_i)_{\Gamma \Omega}\Psi_\Omega \end{equation}

\noindent and where
\begin{equation} 2i L^{-1}(M)\partial_\mu L(M)
= exp({-i}\xi_i A_i)\partial_\mu exp({-i}\xi_j A_j) = \nu_\mu^i
V_i + a_\mu^i A_i \end{equation} have precisely this property.
Secondly, they show that the most general Lagrangian of the type
under consideration maybe written as a function of the standard
fields $\Psi$, $D_\mu \Psi$ and $D_\mu M_i$ only; that is the
$M_i$ will not appear explicitly, and the Goldstone Bosons will be
massless. It then follows that the Lagrangian so formed will be
invariant under $SU_n \times SU_n$ if and only if it is
constructed to be invariant under the $SU_n$ vector subgroup. This
latter requirement is, of course, achieved by index saturation
once more.

The result given in reference [3] (now dropping the chiral
projectors and normalizing for this problem) takes the concrete
form
\begin{equation} D_\mu M_i = \biggl\{ {\partial \psi_\beta \over
\partial\phi_\gamma}(P_{\gamma \beta})_{ik} + {\sum_{A\not= B}}'{\sqrt 2 \over(\phi_A -
\phi_B)}\sin \left[{\psi_A - \psi_B\over \sqrt 2}\right](P_{AB} +
P_{BA})_{ik}\biggr\} (\partial _\mu M_k) \end{equation}

\noindent and represent a complete specification of the required
Lagrangian in simple closed form.  Using the geometric formulation
of Isham [9] gives the coset space metric in the form related to
the covariant derivatives as
\begin{equation} g_{ij}(\partial _\mu M_i)(\partial ^\mu M_j) = (D_\mu
M_i)(D^\mu M_j) \end{equation}

\noindent and we have normalized $g_{ij}$ to $\delta _{ij}$ in the
limit of zero fields.  In matrix notation this yields
\begin{equation} g = {1 \over 4} \left\{p_{\beta \lambda}{\partial \psi_\alpha \over
\partial \phi_\beta}{\partial \psi _\alpha \over \partial \phi _\lambda} + \sum _{A\not= B}
{'} {2\over (\phi_A - \phi_B)^2}(P_{AB} + P_{BA}) \sin^2 \left[
{\psi_A - \psi_B \over \sqrt 2} \right] \right\}
\end{equation}

\noindent immediately because of the orthonormality.

At last it is time to see how this structure is related to $CP_n$.
Returning to the $SU_n \times SU_n$ action given in equations (27)
and (28), consider the restriction of $\xi_i$ to the subset of
dimension $2(n - 1)$ given by
\begin{equation} A_{(n-1)^2},A_{(n-1)^2 + 1},\ldots , A_{n^2 - 2} \end{equation}

\noindent and similarly the restriction of $V_i$ to the subset of
dimension $(n - 1)^2$ given by
\begin{equation} V_1, V_2, \ldots , V_{(n-1)^2 - 1} = V_{n(n-2)}
{\rm \hskip 0.5cm and \hskip 0.5cm} V_{n^2-1} = V_{(n+1)(n+1-2)}
\end{equation}

\noindent which restrictions are overall obviously unique.  The
remaining $V_i$ after the restriction clearly generate $SU_{n-1}
\times U_1$, and the remaining $A_i$ combine with the $V_i$ to
yield the whole $SU_n$ in which the former are uniquely embedded.
This gives the manifold $(SU_n / (SU_{n-1} \times U_1)$ which is
of dimension $2(n - 1)$ and forms the basis for $CP_n$.  All the
previous results now apply to this embedded space simply by
applying the same restrictions.

It is still necessary to interpret the information thus obtained
in terms of the $CP_n$ structure.  From this viewpoint the $V_1,
V_2, \ldots, V_{n^2 - 1}$ and $V_{n(n+2)}$ generate an $SU_{n-1}
\times U_1$ under which the $A_\mu$ transform linearly as a
complex $2(n - 1)$ dimensional multiplet.

Recall that the $2(n - 1) \hskip 0.2cm \xi_i$ are the generalized
coordinates or interpolating fields for the massless Goldstone
Bosons.  We can combine these into $(n - 1)$ complex coordinates
by taking
$$z_0 = \xi_0 - i\xi_1, z_1 = \xi_2 - i\xi_3,\ldots ,z_{n-2} =
\xi_{2(n-1)} - i\xi_{2(n-1) + 1}$$ by a judicious choice of
labels.

\noindent We can see that with the new labels then
$$M =  \sum_{\mu=0}^{n-2} \left[ z_\mu{[\lambda_{2\mu +
(n - 1)^2} + i\lambda_{2\mu + (n - 1)^2 + 1}] \over 2}\right.$$

$$\left. + \bar z_\mu {[\lambda_{2\mu + (n - 1)^2} -
i\lambda_{2\mu + (n - 1)^2 + 1}] \over 2}\right]$$

\noindent having only non-zero entries going from $z_0$ to
$z_{n-2}$ down the final right hand column from the top, and going
from $\bar z_0$ to $\bar z_{n-2}$ across the final row from the
left. There are zeros in the top $(n - 3) \times (n - 3)$ left
hand block, and a zero in the bottom right hand corner.  Each of
the complex $z's$ gives two vectors in the coset space.  The
corresponding lengths can be expressed in terms of the independent
$(n - 2)$ invariants $\phi _\alpha$ out of which the $(n - 2)$
independent functions $\theta _\alpha$ (used in constructing the
$z's$) are formed.
\newline
\hrule width 14cm \vskip 0.3cm \noindent{\bf{The Special Cases of
CP2 and CP4}} \vskip 0.5cm

The CP2 case has previously been called the chiral 2-sphere by
Barnes, Generowicz and Grimshare [11] when it has been described
in some detail.  In the present notation $M$ takes the form
\begin{equation} M = {1 \over 2}{(z + \bar z)}\sigma_1 + {1 \over 2}i{(z - \bar
z)}\sigma_2 = M_A \sigma_A \end{equation} where $z_0$ is written
as $z$ and where $\phi^2 = z \bar z$, and writing $M_A = \phi n_A$
gives
\begin{equation} (P_{12} + P_{21})_{AB} = \delta_{AB} - n_An_B
\end{equation}

\noindent Thus putting
\begin{equation} \psi_1 - \psi_2 = \sqrt 2
\theta \end{equation} and
\begin{equation} \phi_1 - \phi_2 = \sqrt 2 \phi \end{equation}

\noindent one finds immediately that
\begin{equation} g_{AB} = {1 \over 4}\left[\left({d\theta \over d\phi}\right)^2 n_A n_B +
{\sin^2\theta\over\phi^2} (\delta_{AB} - n_A n_B) \right] \hskip
0.5 cm .
\end{equation}

\noindent Note that in this example where there is only a single
arbitrary function $\theta$ of a single invariant $\phi$, the
notation of the $\delta U_2$ description does not need adapting
for the $SU_2/U_1$ coset space.

\noindent The condition to find hermitian form is obviously
\begin{equation} \left({d\theta \over d\phi}\right)^2 = {\sin^2 \theta \over
\phi^2} \end{equation} with the solution
\begin{equation} \phi = ctan\left({\theta \over 2}\right) \end{equation}

\noindent where $c$ is a constant, being the one conventionally
chosen. This is the coordinate system usually known as
stereographic. Obviously equation (71) now yields
\begin{equation} g_{AB} = {{\delta _{AB}}\over [1 + z \bar z]^2}
\end{equation}

\noindent where \begin{equation} z_0 = M_1 + iM_2 {\rm, \hskip 0.2
cm when \hskip 0.2 cm} c = 1
\end{equation}

\noindent and hence it follows that
\begin{equation} {\mathcal L}_2 = {1 \over 2} g_{AB}(\partial_\mu
M_A)(\partial ^\mu M_B) = {1 \over [1 + z_0 \bar z_0]^2}{(\partial
_\mu z_0)(\partial ^\mu \bar z_0) \over 2} \end{equation}

\noindent in obvious hermitian form in these stereographic
coordinates, and sometimes this is written as
\begin{equation} {\mathcal L_2} = {(\partial _\mu \xi)(\partial ^\mu \bar \xi)
\over 2[1 + \xi \bar\xi]^2} \end{equation} where $z_0 = \xi r $ is
used to emphasize the constant radius $r$ of the 2-sphere.

The $CP4$ case has
\begin{equation} M = M_\mu \lambda_\mu = {(z_0 + \bar z_0)\over 2}
\lambda_4 + i{(z_0 - \bar z_0)\over 2} \lambda_5 + {(z_1 + \bar
z_1)\over 2} \lambda_6 + i{(z_2 - \bar z_2)\over 2} \lambda_7
\end{equation}

\noindent This is perhaps a suitable place to note that if the
functions $z_0$ and $z_1$ are not chosen carefully then $M$ will
not be generic and the degree of the equation satisfied by it will
be less than the maximum.

The Goldstones Bosons of this scheme are the octet of pseudo
scalar mesons described by the $M_i$.  In general there are two
$SU_3$ invariants which maybe constructed from the $M^i$.  These
can be denoted
\begin{equation} X = M^iM_i \end{equation}

\noindent and \begin{equation} Y = d_{ijk}M^iM^jM^k \end{equation}

\noindent where the determinantal inequality
\begin{equation} 3Y^2 \leq X^3 \end{equation}

\noindent ensures that the norm of an arbitrary vector constructed
from the $M^i$ shall be positive definite.  Now define $\phi$ and
$\delta$ by
\begin{equation} \phi = X^{1 \over 2} \end{equation}
and
\begin{equation} \phi^3 \sin \delta = \sqrt 3 Y \end{equation}
as the basic invariants.

\noindent It is straightforward to show  [12] that, if
\begin{equation} N_i = d_{ijk}M^jM^k \end{equation}
then
\begin{equation} \hat m_i = \phi^{-1}M_i \end{equation}
and
\begin{equation} \hat r_i = \phi^{-2}\sec \delta(\sqrt 3 N_i - \phi M^i \sin \delta)
\end{equation}

\noindent are an orthonormal base for the independent vectors.

\noindent It has also been shown that the vectors
\begin{equation} q_i = \hat r_i \cos \alpha + \hat m_i \sin\alpha \end{equation}
and
\begin{equation} s_i = (-)\hat r_i \sin \alpha + \hat m_i \cos\alpha
\end{equation}with
\begin{equation} 3\alpha = \delta - 2A\pi \hskip 1 cm (A = 1, 2,
3) \end{equation}

\noindent are respectively charge and special vectors in the sense
of Michel and Radicati [5].  Apart from their orthonormality these
vectors also have the properties
\begin{equation} (-)\sqrt 3 d_{ijk} q^j q^k = q_i = \sqrt 3 d_{ijk} s^j s^k
\end{equation}
\begin{equation} \sqrt 3 d_{ijk} s^j q^k = s_i \end{equation}
and \begin{equation} f_{ijk} s^j q^k = 0 \end{equation}

\noindent so that a single pair $q^i$ and $s^i$ represent a useful
alternative to working with the three $P^A_i$ which are linearly
dependent.  Adopting the choice $A = 3$ for the set of standard
$q^i$ and $s^i$ it follows that
\begin{equation} \sqrt 2 p_1^i = q_i + s_i \end{equation}
and \begin{equation} \sqrt 2 p_2^i = q_i - s_i \end{equation} are
the orthonormal basis vectors introduced previously.

The second rank tensors which maybe constructed from the $M^i$ are
spanned by the six projection operators $(P_{AB})_{ij}$ and the
four $(p_{\alpha\beta})_{ij}$, all of which are taken to be
hermitian in the matrix sense.

It is standard to introduce projection operators with a cyclic
notation in the form
\begin{equation} (S_1)_{ij} = (P_{23})_{ij} + (P_{32})_{ij} \end{equation}
\begin{equation} (S_2)_{ij} = (P_{13})_{ij} + (P_{31})_{ij} \end{equation}
\begin{equation} (S_3)_{ij} = (P_{12})_{ij} + (P_{21})_{ij} \end{equation}

\noindent The first term in equation (64) can be treated by making
the substitutions (where lower and upper Greek indices take the
ranges 4-5 and 6-7 respectively)
\begin{equation} p_1 ^\mu \Rightarrow n_1^\mu ,\hskip 0.4cm p_2^\Gamma \Rightarrow n_2^\Gamma 
\end{equation}

\begin{equation} {\partial \psi^\beta \over \partial_ \mu \phi_1}
\Rightarrow \sqrt 2 {\partial \psi^ \beta \over \partial _\mu
\omega_1} = \sqrt 2 {\partial \psi^ \beta \over \partial \omega_1}
{\partial \omega_1 \over \partial M_\mu} = \sqrt 2 {\partial \psi^
\beta \over \partial \omega_1} {M_\mu \over \omega _1}
\end{equation}

\begin{equation} {\partial \psi^\beta \over \partial_ \Gamma \phi_2}
\Rightarrow \sqrt 2 {\partial \psi^ \beta \over \partial _\Gamma
\omega_2} = \sqrt 2 {\partial \psi^ \beta \over \partial \omega_2}
{\partial \omega_2 \over \partial M_\Gamma} = \sqrt 2 {\partial
\psi^ \beta \over \partial \omega_2} {M_\Gamma \over \omega _2}
\end{equation}

and similarly, using equations (95), (96) and (97), the $S_A$ can
be brought to the forms
\begin{equation} (S_1)_{\mu \upsilon} = \delta_{\mu \upsilon} -
n'_\mu n'_\upsilon \end{equation}
\begin{equation} (S_2)_{\Gamma \Omega} = \delta_{\Gamma \Omega} -
n^2_\Gamma n^2_\Omega \end{equation}

\noindent and $S_3$ vanishes because $n^3$ lies inside the $SU_2
\times U_1$ subspace rather than in the coset space.  [This
explains why the range of summation is reduced in future.]

\noindent It follows that
\begin{equation} 4g_{\mu \upsilon} = 2{\partial \psi ^\beta \over \partial \omega_1}
{\partial \psi ^\beta \over \partial \omega_1}{M_\mu M_\upsilon
\over (\omega_1)^2} + {4(S_1)_{\mu \upsilon}\over (\omega_1)^2}
\sin^2 \left({\theta_1 + 2\theta_2 \over 2}\right) \end{equation}

\begin{equation} 4g_{\Gamma \Omega} = 2{\partial \psi ^\beta \over \partial \omega_2}
{\partial \psi ^\beta \over \partial \omega_2}{M_\Gamma M_\Omega
\over (\omega_1)^2} + {4(S_2)_{\Gamma \Omega}\over (\omega_2)^2}
\sin^2 \left({\theta_2 - 2\theta_1 \over 2}\right)
\end{equation}

\noindent and
\begin{equation} 4g_{\mu \Omega} = 2{\partial \psi ^\beta \over \partial \omega_1}
{\partial \psi ^\beta \over \partial \omega_2}{M_\mu M_\Omega
\over \omega_1 \omega_2} \end{equation}

\noindent Noting that $M_\mu \equiv \omega_1 n^1_\mu $, it appears
that the hermiticity conditions on the diagonal components are
\begin{equation} {\partial \psi ^\beta \over \partial \omega_1}
{\partial \psi ^\beta \over \partial \omega_1} = {2 \over (\omega
_1)^2} \sin ^2 \left({\theta_1 + 2\theta _2 \over 2} \right)
\end{equation}
and
\begin{equation} {\partial \psi ^\beta \over \partial \omega_2}
{\partial \psi ^\beta \over \partial \omega_2} = {2 \over (\omega
_2)^2} \sin ^2 \left({\theta_2 - 2\theta _1 \over 2} \right)
\hskip 0.5cm .
\end{equation}
These conditions can be imposed by a slight generalisation of the
method used in the CP2 case.  Obviously it will be advantageous to
introduce the abbreviations
\begin{equation} D = {\partial \psi_1 \over \partial \omega _1}
\hskip 0.3 cm, \hskip 0.3 cm D' = {\partial \psi_2 \over \partial
\omega _1} \hskip 0.3 cm, \hskip 0.3 cm d = {\partial \psi_1 \over
\partial \omega _2} \hskip 0.3 cm \rm {and} \hskip 0.3 cm
d' = {\partial \psi_2 \over \partial \omega _2} \end{equation}

\noindent It is simple to see that from equations (106) and (107)
it follows that
\begin{equation} D^2 + D'^2 = {2 \over (\omega
_1)^2} \sin ^2 \left[{\theta_1 + 2\theta _2 \over 2} \right]
\end{equation}
and
\begin{equation} d^2 + d'^2 = {2 \over (\omega
_2)^2} \sin ^2 \left[{\theta_2 - 2\theta _1 \over 2} \right]
\end{equation}

\noindent These two results ensure the hermiticity constraints on
$g_{\mu \nu}$ and $g_{\Gamma \Omega}$, which then takes forms

\begin{equation} 4g_{\mu \upsilon} = {1 \over (\omega
_1)^2} \delta_{\mu\upsilon} sin ^2 \left({\theta_1 + 2\theta _2
\over 2} \right) \end{equation} and
\begin{equation} 4g_{\Gamma \Omega} = {1 \over (\omega
_2)^2} \delta _{\Gamma\Omega}\sin^2 \left({\theta_2 - 2\theta _1
\over 2} \right) \end{equation}

\noindent where the normalization has again been adjusted.  The
other components take the forms

\begin{equation} 4g_{\mu \Omega} = (Dd + D'd'){n^1_\mu n^2_\Omega \over 2}
\end{equation}
and
\begin{equation} 4g_{\Gamma \upsilon} = (Dd + D'd'){n^1_\upsilon n^2_\Gamma \over 2}
\end{equation}
and as these are off diagonal it is necessary to show hermiticity
makes them zero.

\noindent Now put
\begin{equation} \omega_1 = [c^2 + (\omega_2)^2]^{1 \over 2} \tan
\left({\theta_1 + 2 \theta_2 \over 4} \right) \end{equation}

\noindent to show that
\begin{equation} {\partial \left({\theta_1 + 2 \theta_2
\over 2} \right)\over \partial \omega_1} = {\sin \left({\theta_1 +
2 \theta_2 \over 2}\right) \over \omega_1} = {2[c^2 +
(\omega_2)^2]^{1 \over 2}\over [c^2 + (\omega_1)^2 + (\omega_2)^2]
} \hskip 0.5cm .\end{equation}

\noindent Similarly, put
\begin{equation} \omega _2 = [c'^2 + (\omega_1)^2]^{1 \over 2}
\tan \left( {\theta_2 - 2 \theta_1 \over 4} \right)\end{equation}
to show that
\begin{equation} {\partial \left({\theta_2 - 2 \theta_1
\over 2} \right)\over \partial \omega_2} = {\sin \left({\theta_2 -
2 \theta_1 \over 2}\right) \over \omega_2} = {2[c'^2 +
(\omega_1)^2]^{1 \over 2}\over [c'^2 + (\omega_1)^2 + (\omega_2)^2
]}\hskip 0.5cm .\end{equation}

\noindent It is straightforward to see from equations (106) and
(107) that
\begin{equation} \left({\partial \theta _1 \over \partial
\omega_1}\right)^2 + \left({\partial \theta _1 \over \partial
\omega_1}\right) \left({\partial \theta _2 \over \partial
\omega_1}\right) + \left({\partial \theta _2 \over \partial
\omega_1}\right)^2 = {1 \over (\omega_1)^2}\sin^2 \left({\theta _1
+ 2\theta _2 \over 2}\right)
\end{equation}

\noindent and
\begin{equation} \left({\partial \theta _1 \over \partial
\omega_2}\right)^2 + \left({\partial \theta _1 \over \partial
\omega_2}\right) \left({\partial \theta _2 \over \partial
\omega_2}\right) + \left({\partial \theta _2 \over \partial
\omega_2}\right)^2 = {1 \over (\omega_2)^2}\sin^2 \left({\theta _2
- 2\theta _1 \over 2}\right) \end{equation} so that, using the
left hand parts of equations (116) and (118), it follows that
\begin{equation} c' = (\pm)c \end{equation}

is required since the $c$ and $c'$ are constants independent of
the $\omega_1$ and $\omega_2$ variables, and the expressions on
the right hand sides result simply from using a trigonometric
substitution to evaluate the integral.  Hence, the forms
\begin{equation} g_{\mu \upsilon} = {\delta_{\mu \upsilon}[c^2 +
(\omega_2)^2]\over [c^2 + (\omega_1)^2 + (\omega_2)^2]^2}
\end{equation}
and \begin{equation} g_{\Gamma \Omega} = {\delta_{\Gamma
\Omega}[c^2 + (\omega_1)^2]\over [c^2 + (\omega_1)^2 +
(\omega_2)^2]^2} \end{equation} are revealed. \newline
\newline
Now consider
\begin{equation}D^2 + D'^2 = d^2 + d'^2 \end{equation}

\noindent which follows from equations (109) and (110) by using
the left hand parts of equations (116) and (118), now that $c' =
(\pm)c$, reveals that
\begin{equation} dD + d'D' = 0 \hskip 0.5cm .\end{equation}

\noindent From equations (113) and (114), it is now evident that
\begin{equation} g_{\mu \Omega} = 0 = g_{\Gamma \upsilon} \end{equation}

\noindent and this completes the specification of the metric
through the hermiticity conditions.  It is perhaps worth repeating
that the forms of the metric (given for example in equation (103)
and (104)) are in general coordinates before the hermiticity
conditions are applied.  However, the forms given in equations
(122) and (123) are in hermitian form and maybe directly compared
with the classic results of Fubini [13] and Study [14].  These
authors apply scaling by using the conformal symmetry of the
metric, and this is directly equivalent to setting $c = 1$ in the
present notation.  Reverting to complex notation reveals the
invariant
\begin{equation} {\mathcal L_4} = {dz_1 d\bar z_1[1 + z_2 \bar z_2] + dz_2 d\bar z_2
[1 + z_1 \bar z_1] \over [1 + z_1 \bar z_1 + z_2 \bar z_2]^2}
\end{equation}

\noindent retrieving the Fubini-Study form.\newline
\newline
\noindent {\bf Conclusions}\newline It appears that the CPN metric
has been found in general coordinates (in principle) for all N,
and that in the cases of $N = 2$ and $N = 4$ well known forms are
recovered in the hermitian limit.  Obviously, the algebraic effort
required does rise with $N$ but only simple functions ever
appear.\newline
\newline
\noindent {\bf Acknowledgements}\newline

This work is supported in part by PPARC grant number GR/L56329.
The author is grateful to Dr. J. A. Vickers for help and advice on
Kahler manifolds.  Particular gratitude must be offered to
Professor W. J. Zakrzewski who performed the miracle of persuading
the author that not only was this problem important but that he
had an outside chance of solving it.\newline \hrule width 14 cm

\end{document}